\documentclass[%
 notitlepage,
 reprint,
 superscriptaddress,
 amsmath,amssymb,
 aps,
]{revtex4-1}

\usepackage{xcolor}
\usepackage{lipsum}

\usepackage{graphicx}
\usepackage{dcolumn}
\usepackage{bm}
\usepackage{hyperref}

\usepackage{epstopdf}

\newcommand{\bra}[1]{\langle #1 \vert}
\newcommand{\ket}[1]{\vert #1 \rangle}

\begin{document}

\preprint{APS/123-QED}

\title{Rogue waves in discrete-time quantum walks}

\author{A. R. C. Buarque, W. S. Dias, F. A. B. F. de Moura, M. L. Lyra, G. M. A. Almeida}
\affiliation{
Instituto de F\'isica, Universidade Federal de Alagoas, 57072-900 Macei\' o, Alagoas, Brazil
}%
\begin{abstract}
Rogue waves are rapid and unpredictable events of exceptional amplitude reported 
in various fields, such as oceanography and optics, with much
of the interest being targeted towards their 
physical origins and likelihood of occurrence.
Here, we use the all-round framework of discrete-time quantum walks to study the onset of those events 
due to a random phase modulation, unveiling its long-tailed statistics, distribution profile, and dependence upon the degree of randomness. We find that those rogue waves belong the Gumbel family 
of extreme value distributions. 
\end{abstract}
\maketitle

\textit{Introduction}. --- Rogue or freak waves, unpredictable and rare huge walls of water
appearing from nowhere and vanishing without a trace, 
have been known and feared for centuries by seafarers. 
The first solid account of the phenomenon took place 
in 1995 when data collected on the Draupner oil platform in the North Sea 
revealed a 26-meter wave rising out of a background with about half
significant wave height \cite{Harvey2004}. 
Years later, analogies
between such ocean wave phenomena and light propagation in optical fibers surged  
in the framework of the nonlinear Schroedinger equation \cite{solli07}. 
Since then, interest in ubiquitous wave phenomena displaying long-tailed statistics, when outliers occur more often than expected from Gaussian statistics,
has skyrocketed in various fields (for a recent review, see \cite{dudley19}).
%
Optics, particularly, has been a powerful testbed for investigating rogue waves thanks to
the spatial and timescales involved and, 
in addition, optical rogue waves
include a bunch of novel phenomena, not necessarily featuring a hydrodynamics counterpart \cite{dudley19}.

One of the key challenges in the field is to find out precisely how those events emerge
so as to be able to predict and control them. 
There is a long-standing debate on whether rogue waves are primarily driven by 
linear or nonlinear processes \cite{dematteis19} and what is the role of noise and randomness \cite{bonatto11}. It is natural to assume
that nonlinearity plays an important role due to modulational instability \cite{baronio14,xu20}, collisions between solitons \cite{akhmediev09}, and so forth.
On the other hand, some studies suggest that linear interference of random fields are crucial \cite{hohmann10, arecchi11, metzger14, mathis15, liu15, mattheakis16, derevyanko18, peysokhan19, rivas20, bonatto20, frostig20, dasilva20}, with nonlinear effects
responsible for extra wave focusing \cite{onorato01, ying11,safari17}. 
Indeed, linear models can display rogue waves on their own when augmented with the right ingredients as shown in \cite{arecchi11}. 
This has been shown experimentally in microwave transport in randomly distributed scatterers \cite{hohmann10}, 
2D photonic crystal resonators \cite{liu15}, and very recently by measuring linear light diffraction patterns in the presence of long-range spatial memory effects in the random input \cite{bonatto20}. 

Interest on linear rogue waves has been raising considerably over the past few years. Yet, it is surprising that 
quantum mechanics has barely been taken into consideration. 
Even though the dynamics of a single quantum particle can be mapped into linear optics, 
investigating the onset of rogue-like events in the very domain of quantum mechanics
has its own appeal. It could, for instance, shed new light
on the dynamics of disordered systems and related
features such as Anderson localization.
With that in mind, we set about 
to explore the occurrence of \textit{rogue quantum amplitudes} using 
the discrete-time quantum walk (DTQW) approach \cite{kemperev}. It is basically a cellular automaton \cite{meyer97} whose updating rules are run by a preset sequence of quantum gates.
Given recent experimental advances in the field \cite{su19,wang19,alderete20} as well as their
wide range of applications, from quantum algorithms \cite{lovett10} to simulation of involved phenomena in condensed matter physics
\cite{obuse15,rakovszky15,derevyanko18,mendes19,buarque20,mendonca20}, 
DTQWs make for a suitable starting point.
%
%

We report the manifestation of rogue waves in the Hadamard one-dimensional DTQW induced by
random phase fluctuations.
We do so by unveiling the long-tailed statistics of the occupation probability amplitudes (which is analogous to light intensity in optics) over the space-time set of events. We show that an intermediate level of disorder 
scaling as $N^{-\nu}$
maximizes the likelihood of rogue events. That has to do with a fair balance between localization and mobility, for which the localization length $\propto N^{2\nu}$,
$N$ being the number of sites. 
Furthermore, extreme-value analysis is carried out for
the amplitude block maximum over time and we find that the resulting distribution falls into the Gumbel class.
%

\textit{Quantum walk model}. --- We consider a single-particle DTQW in one dimension \cite{kemperev} defined by a two-level (coin) space $H_{C}=\{\ket{\uparrow},\ket{\downarrow}\}$ and a position space $H_{P}=\{\ket{n}\}$, such that the full Hilbert space reads $H=H_{C}\otimes H_{P}$.
An arbitrary state at a given instant $t$ can be written as 
$
\ket{\Psi_{n}(t)} = \sum_{n=1}^{N}
( a_{n}(t) \ket{\uparrow, n}+b_{n}(t) \ket{\downarrow, n} ),
$
satisfying the normalization condition $\sum_{n}P_{n}(t) = \sum_{n}(|a_{n}(t)|^{2}+|b_{n}(t)|^{2})=1$.

The quantum walker evolves as $|\Psi(t+1)\rangle=\hat{S}(\hat{C}\otimes I_{p})\hat{D}|\Psi(t)\rangle$,
where the conditional shift operator $\hat{S}$ is responsible for the nearest-neighbor transitions
$\hat{S}\ket{\uparrow, n}=\ket{\uparrow, n+1}$ and $\hat{S}\ket{\downarrow, n}=\ket{\downarrow, n-1}$ (assuming periodic boundary conditions), 
$\hat{C}=\big(\ket{\uparrow}\bra{\uparrow} + \ket{\uparrow}\bra{\downarrow} + \ket{\downarrow}\bra{\uparrow}
- \ket{\downarrow}\bra{\downarrow}\big)/\sqrt{2}$ is the standard Hadamard coin,
$I_{p}$ is the identity operator
acting on the $N-$dimensional position space, and 
\begin{eqnarray}\label{phase}
\hat{D} &=&\sum_{c}\sum_{n} e^{iF(c,n,t)}|c,n\rangle\langle c,n|
\end{eqnarray}
is the phase-gain operator, with $F(c,n,t)$ being a real-valued arbitrary function \cite{navarrete07} and $c=\uparrow,\downarrow$.
Given the flexibility in choosing $F(c,n,t)$, one is able to produce various dynamical regimes.  
Setting $F=0$ renders the standard Hadamard quantum walk in which walker spreads out ballistically \cite{kemperev}. Here, instead, we set a static random phase modulation such that $F(c,n,t) = F(c,n) =2\pi\nu$, where $\nu$ is a random number uniformly distributed within $[-W,W]$, with $W$ being the disorder width. As this setting can lead to Anderson localization \cite{derevyanko18,mendes19}, we ought to inquire whether rogue waves can be supported given proper initial conditions and amount of noise embedded in $F(c,n)$.

   
\begin{figure}[t!]
\centering 
\includegraphics[width=\linewidth]{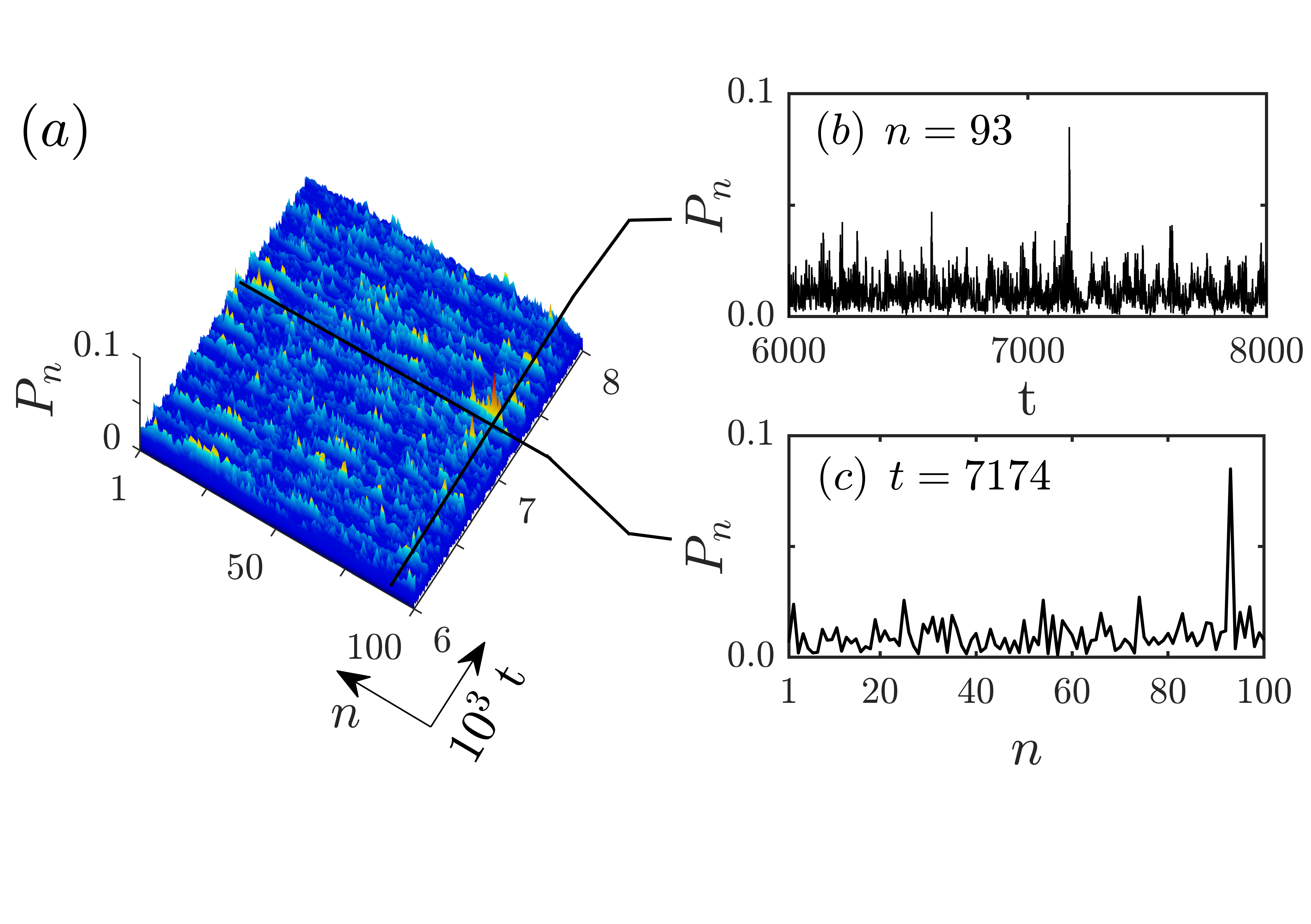}
        \caption{(a) Snapshot of the space-time evolution of the occupation amplitude $P_{n}$ 
in the Hadamard DTQW on a ring with $N=100$ sites and disorder strength $W=0.1$ (single realization). 
(b,c) Time series and spatial profile extracted from (a). The rogue event is seen at $t=7174$.
       }
        \label{fig1}
\end{figure}   
    
\textit{Results}. --- In order to avoid ambiguity between an actual rogue event (a rare one) and the inevitable Anderson localization in the statistics, 
we initialize the system in a coin-unbiased \cite{tregenna03}, fully delocalized state
$|\psi(0)\rangle=\frac{1}{\sqrt{2N}}\sum_{n=1}^{N}(\ket{\uparrow, n}+i\ket{\downarrow, n})$.
Random phase modulation is introduced
at the very first gate operation $\hat{D}$ [see Eq. (\ref{phase})]
so as to foster inhomogeneity 
and, as a result, fragmentation of the walker wavefunction. These two ingredients have been proved to be crucial for the development of linear rogue waves \cite{arecchi11}. 
%
%

%

 
Let us now establish the criteria to identify the rogue waves. A standard approach in oceanography and optics \cite{dudley19}
sets that the amplitude of a rogue wave must exceed at least twice the significant wave height, defined as the mean the largest one third
of a record \cite{kharif_book}. 
%
%
Following that, we define a probability threshold $P_{th}$ evaluated over the whole
space-time evolution for a given realization of disorder so that a rogue event is counted whenever $P_{n}(t)>2P_{th}$. 

Figure \ref{fig1} shows a snapshot of a typical rogue wave event with $P_{n} \approx 5P_{th}$ 
alongside a detailed look over the amplitude record over space and time. The peak
shares all the standard characteristics of a rogue wave: besides the large amplitude
in comparison to the background, it is unpredictable and short-lived.  
%

\begin{figure}[t!]
        \centering 
        \includegraphics[width=0.65\linewidth]{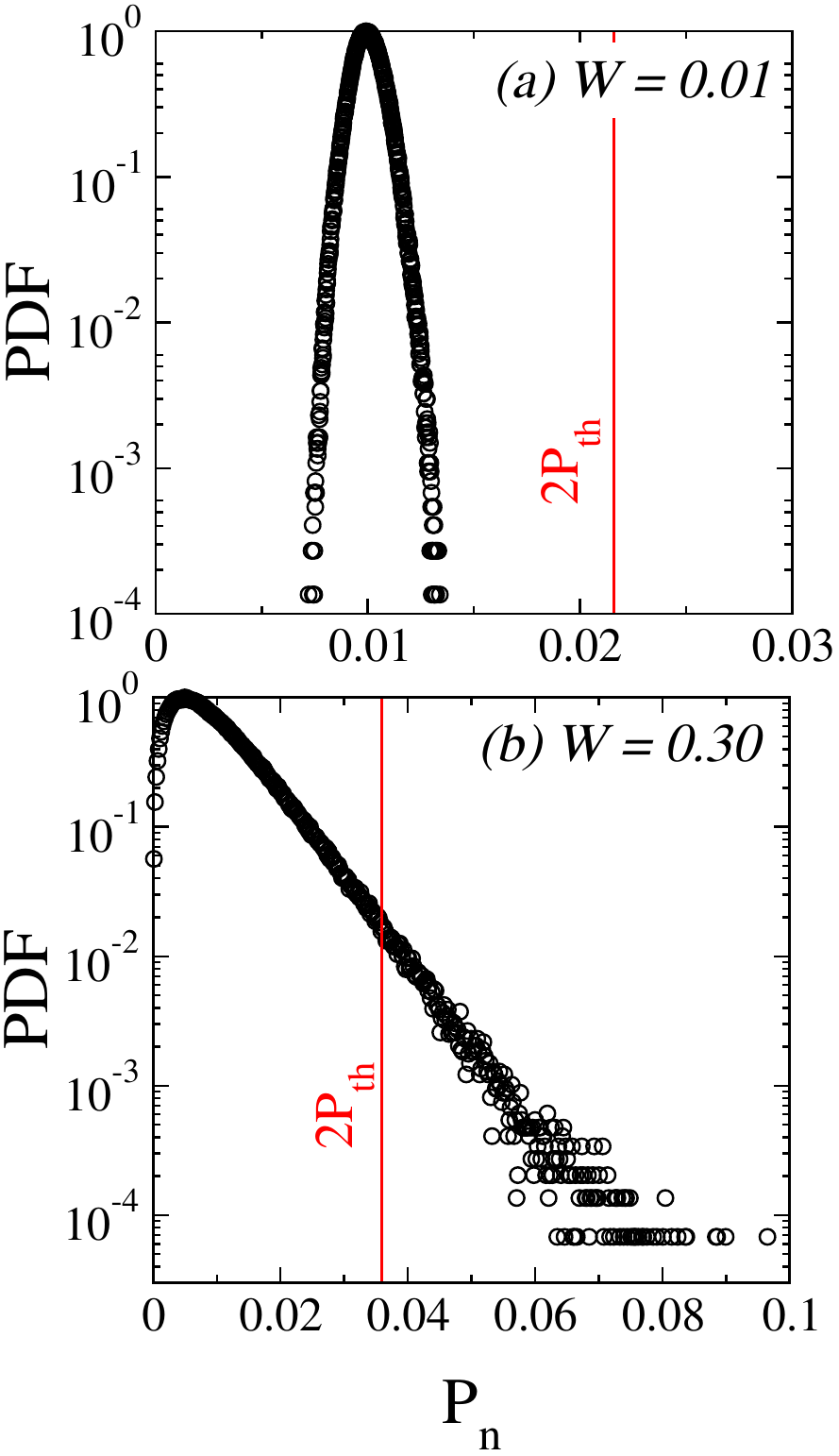}
                \caption{Normalized PDFs for (a) $W=0.01$ and (b) $W=0.3$ in semilog scale for an ensemble of $5000$ independent realizations of disorder and $10^{4}$ steps on a cycle with $N=100$ sites.}
                \label{fig2}
\end{figure}


In Fig. \ref{fig2} we plot some PDFs over the entire ensemble for some representative strengths of disorder,
weak and intermediate.
The later clearly displays another key signature of occurrence of rogue events \cite{dudley19}, 
which is a positively skewed, $L-$shaped distribution. It features a significant number of outliers in the high-amplitude
range, relatively rare among the total number of events yet more than what one would get from Gaussian statistics. 

\begin{figure}[!t]
        \centering 
        \includegraphics[width=0.85\linewidth]{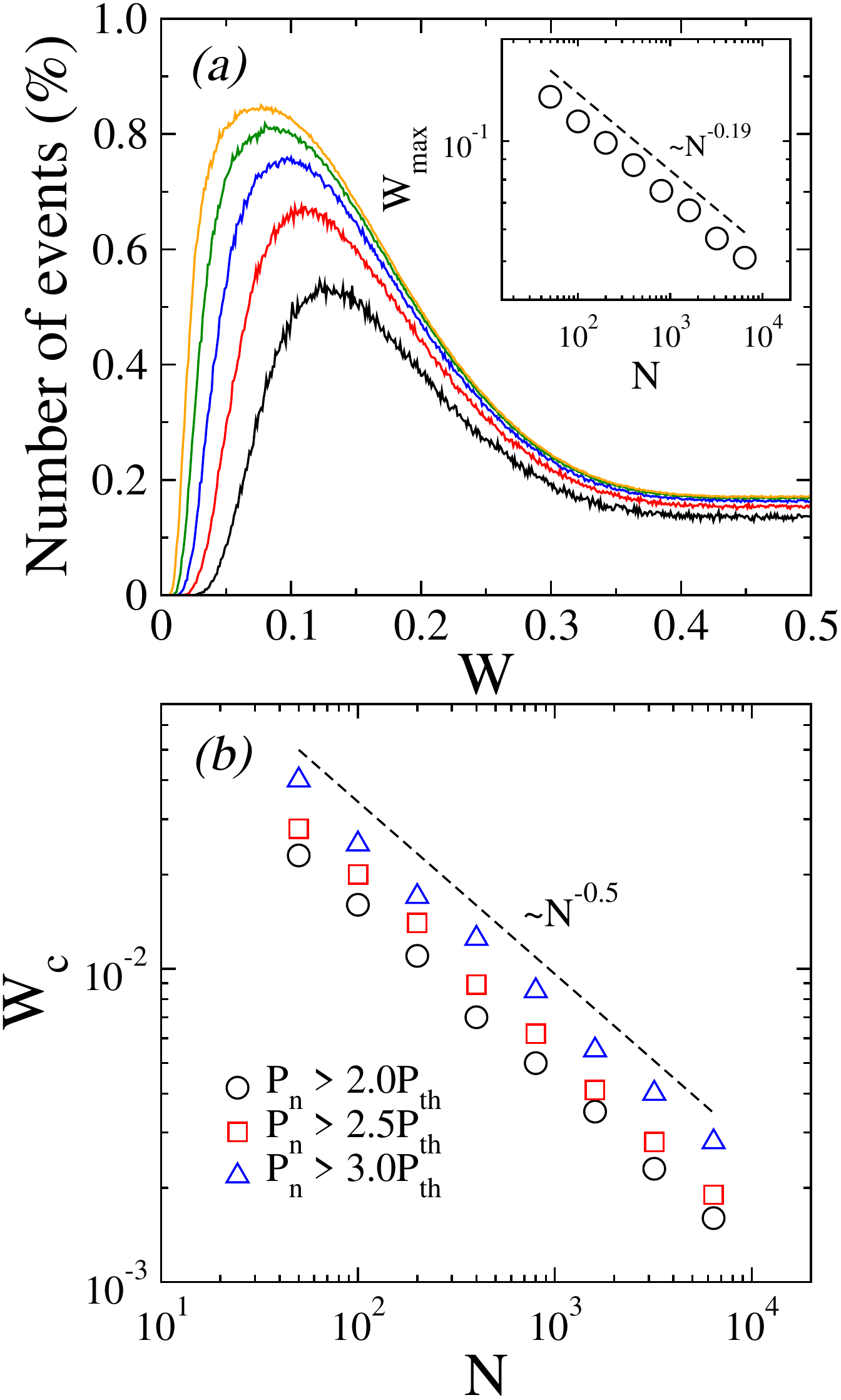}
                \caption{(a) Number of rogue wave events versus disorder strength $W$ for $N=50, 100, 200, 400$ and $800$ sites (black line to the orange, respectively),                  
averaged over 5000 independent realizations of disorder, each running through $10N$ times steps. Inset shows the scaling of 
the disorder degree that maximizes
the chances of measuring a rogue event, $W_{\mathrm{max}}$, with $N$. (b) Disorder strength $W_c$ above which rogue waves have a finite occurrence probability for distinct threshold levels. The scaling $W_c\propto N^{-1/2}$ unveils that at $W_c$ the localization length $\chi\propto 1/W^2$ is of the order of the chain size.  
               }
                \label{fig3}
\end{figure}

\begin{figure}[!t] 
        \centering 
        \includegraphics[width=0.85\linewidth]{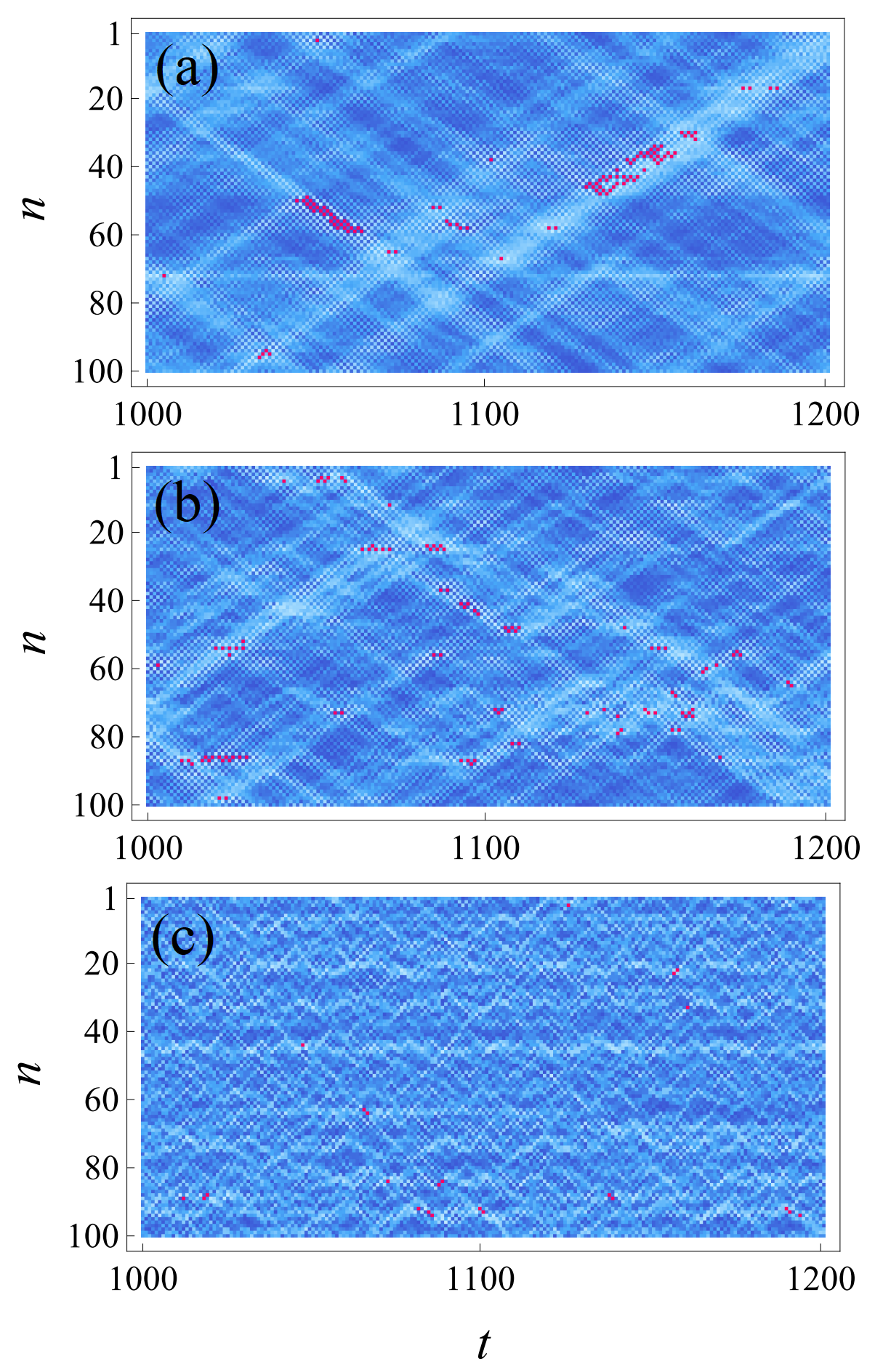}
                \caption{Space-time branching patterns of $P_{n}$ for disorder widths (a) $W=0.05$,
                (b) $W=0.1$, and (c) $W=0.5$ (single realizations). Red spots (see online version) are rogue wave events fulfilling $P_{n}>2P_{th}$ for each sample.
               }
                \label{fig4}
\end{figure}

\begin{figure}[!t]
        \centering 
        \includegraphics[width=0.65\linewidth]{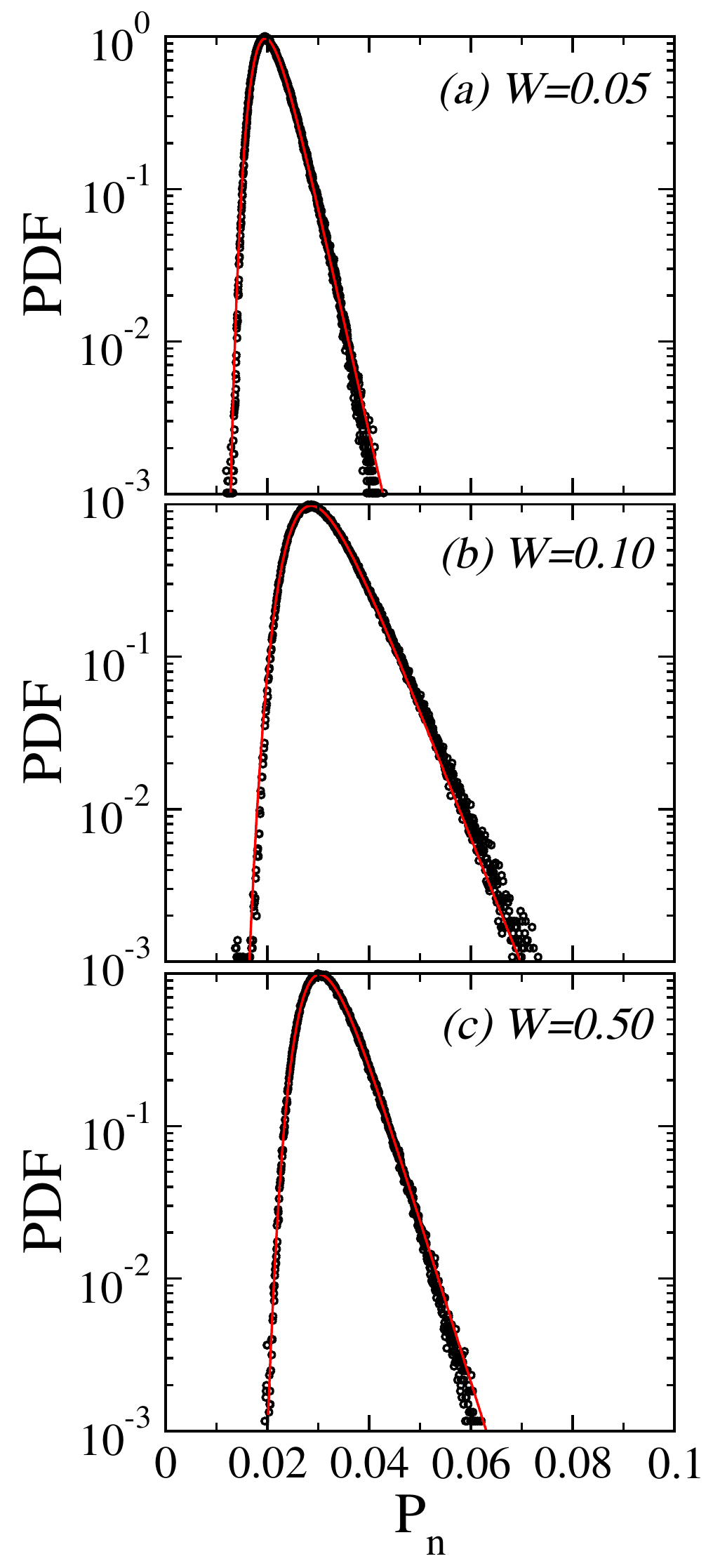}
                \caption{Extreme-value PDFs in semilog scale for $N=100$ and $10^{4}$ independent realizations of disorder.
At each time step, the maximum probability amplitude $P_{\mathrm{max}}$ is recorded.                
                 The red line is a Gumbel-type fitting given by $y(x)\propto \exp[-ax-b\exp(-ax)]$, with $(a,b)$ being $(333.52, 396.98)$ for (a) $W=0.05$,
$(184.73, 106.12)$ for                
(b)$W=0.10$, and (232.51, 1002.07) for  (c) $W=0.5$.
               }
                \label{fig5}
\end{figure}

In order to analyze those distributions in a more quantitative level for the whole range of $W$, Fig. \ref{fig3}(a) shows the
ensemble-averaged percentage of events fulfilling
$P_{n}>2P_{th}$.
Fig \ref{fig3}(b) shows the size dependence of the disorder level $W_c$ above which extreme events have a finite occurrence probability. Irrespective of the threshold level, the minimal disorder strength leading to the occurrence of rogue waves $W_c\propto N^{-1/2}$. It is interesting to stress that the typical localization length of the eigenstates of quantum walks under random phase shifts scales as $\chi\propto 1/W^2$ \cite{derevyanko18}. Therefore,  $\chi\propto N$ at $W_c$. The above result unveils that rogue waves emerge whenever disorder is strong enough to produce effectively localized states in a finite chain with $N$ sites.  Curiously, there is an optimal
level of disorder, $W_{\mathrm{max}}$, that maximizes
the chances of observing a rogue event somewhere along the
$N$-site cycle. This suggests that rogue events are more likely to develop when one properly 
balances localization and mobility.
%
The inset of Fig. \ref{fig3}(a) 
shows that $W_{\mathrm{max}}\propto N^{-\nu}$, with $\nu\approx 0.19$ over the range of chain sizes considered. For such disorder level, the typical localization length scales as $\chi\propto N^{2\nu}$.
%



%


An increased likelihood of the occurrence of rogue waves between weak and intermediate disorder strengths has been seen in recent
experiments carried out on 1D photonic lattices featuring both on-site and coupling disorder \cite{rivas20}.
That also suggests that the interplay between localization and delocalization is a key ingredient for the for the generation of extreme events in linear systems. 
Furthermore, correlated fluctuations -- known to yield rich transport properties \cite{izrailev_rev} -- have been exploited to enhance 
the likelihood of occurrence of rogue waves \cite{peysokhan19,bonatto20},
some of these largely exceeding the amplitude threshold (refereed to as super rogue waves) \cite{bonatto20}. 

Large fluctuations in $F(c,n)$ [cf. Eq. (\ref{phase})] tend to make localization effects sharper but it does not necessarily mean that
the occurrence of rogue waves will follow that up. 
We shall always keep in mind that a rogue wave is a rare and sudden event whose amplitude 
should exceed some threshold based on the average amplitude background. In order to produce 
such abnormal constructive interference at some location via linear dynamics, we need
proper synchronization of random waves undergoing 
different paths and thereby some degree of mobility. 
%
Figure \ref{fig4} shows
the evolution of branching patterns highlighting the distribution profile of the rogue events (red spots). 
In the case of weak disorder, we note that whenever synchronization
conditions are met to form a rogue wave, it usually covers a few sites in the neighborhood before disappearing [\ref{fig4}(a)]. 
For intermediate disorder, the rogue events become sparse but more frequent, as a more complex
branching profile emerges [\ref{fig4}(b)].  
If we keep on increasing the disorder width $W$, there will be a stage above which 
mobility, if any, is restricted to shorter spatial domains given
the onset of local resonances. 
This is seen in Fig. \ref{fig4}(c) in the form of well defined
amplitude domains, with few of them giving rise to rogue waves now and then. That is why the rogue-wave likelihood saturates for large $W$ and barely responds to the system size $N$ [see Fig. \ref{fig3}(a)].


%
%




Last but not least, we carry out an extreme value analysis
by selecting the maximum amplitude at each time step during the evolution. We do so for many samples so as to 
generate another distribution and see whether it
falls into one the three limiting types, namely Weibull,
Fr\'{e}chet, or Gumbel,
according to a general theorem in
extreme value theory \cite{laurens}. Figure \ref{fig5} shows that
our extreme events belong to the Gumbel class, suited
by a distribution of the form
$y(x)\propto \exp[-ax-b\exp(-ax)]$, with $a,b$, depending on
$W$ and $N$. For intermediate degree of disorder, 
as in Fig. \ref{fig5}(b), the range of $P_{\mathrm{max}}$ is visibly
more stretched, what again indicates a pronounced likelihood of observing a rogue event.

\textit{Final remarks}. --- We have reported 
the occurrence of rogue wave events in disordered DTQWs and showed that those indeed belong a class of extreme value phenomena. Using the peak-over-threshold approach borrowed from oceanography, we have also uncovered the long-tailed profile of the distributions.
We found that an intermediate degree of disorder $W_{\mathrm{max}}\propto N^{-\nu}$ yields maximum occurrence of rogue waves 
due to a proper balance between trapping mechanisms and mobility for which $\chi \propto N^{2\nu}$. This calls for further investigation in order to assess the intrinsic relationship between localization length and rogue wave generation, specially in the case of correlated phases which has been shown to enhance the occurrence of  extreme events \cite{peysokhan19, bonatto20}. 

The DTQW studied here also offers the possibility of embedding nonlinearity into $F(c,n,t)$. In \cite{navarrete07}, for instance,
the authors considered a Kerr-type self-phase modulation and reported the formation of solitonlike pulses. Also, in \cite{buarque20}, it was shown that self-trapping can occur for certain coin angles.   
The stage is thus set for assessing the competition between linear and nonlinear mechanics in the generation of rogue waves in quantum walks. 

We hope that our work seeds interest in quantum-mechanical extreme events in general, specially in 
the context of condensed-matter theory, for the sake of
facing Anderson localization phenomena under different light,  
as well as in the field of quantum information processing, where unexpected events of that nature could lead to potential hazards in the evaluation of some protocol given the unavoidable presence of manufacturing imperfections of the physical components. 



This work was supported by CNPq, CAPES (Brazilian Federal Agencies), and FAPEAL (Alagoas State Agency).

%

\end{document}